\documentclass{camera}
\usepackage{graphicx}  
\usepackage{subfigure}
\begin{document}
%
\title{Light fragments from (C + Be) interactions at 0.6 GeV/nucleon}
%
\author{
B.M. Abramov$^{1,2}$, P.N. Alekseev$^{1}$, Yu.A. Borodin$^{1}$,
S.A. Bulychjov$^{1}$, I.A. Dukhovskoy$^{1}$, A.I. Khanov$^{1}$,
A.P. Krutenkova$^{1}$, V.V. Kulikov$^{1}$, M.A. Martemianov$^{1}$,
S.G. Mashnik$^{3}$, M.A. Matsyuk$^{1}$, E.N. Turdakina$^{1}$
\and P.I. Zarubin$^{4}$}
\organization{$^{1}$Institute for Theoretical and Experimental
Physics (ITEP) SRC "Kurchatov institute", Moscow, 117218, Russia,\\
$^{2}$Moscow Institute of Physics and Technology (MIPT), Dolgoprudnyy,
117303, Russia, \\$^{3}$Los Alamos National Laboratory (LANL), Los
Alamos, NM, 87545, USA \and \\$^{4}$Joint Institute for Nuclear
Research, Dubna, 141980, Russia}
\maketitle
\begin{abstract}
Nuclear fragments emitted at 3.5$^{\circ}$ in $^{12}$C fragmentation at
0.6 GeV/nucleon have been measured. The spectra obtained are used for
testing the predictions of four ion-ion interaction models:
INCL++, BC, LAQGSM03.03 and QMD as well as for the comparison with
the analytical parametrization in the framework of
thermodynamical picture of fragmentation.
\end{abstract}
\section{Introduction}
\label{intro}
The study of emission of light fragments is important to understand
the nature of ion-ion interactions. Different reaction mechanisms
contribute to this rather complicated process which can hardly be
described in analytical way. For this reason we tested a few
Monte-Carlo transport codes against the data of the FRAGM experiment \cite{jetpl,epj,Yaf2015}.
\section{ The FRAGM Experiment and the test of the models of ion-ion interactions}
\begin{figure}[ht!]
     \begin{center}
        \subfigure[]{%
    \label{600MeVINCL}\hspace{-0.3cm}
          \includegraphics[width=0.50\textwidth]{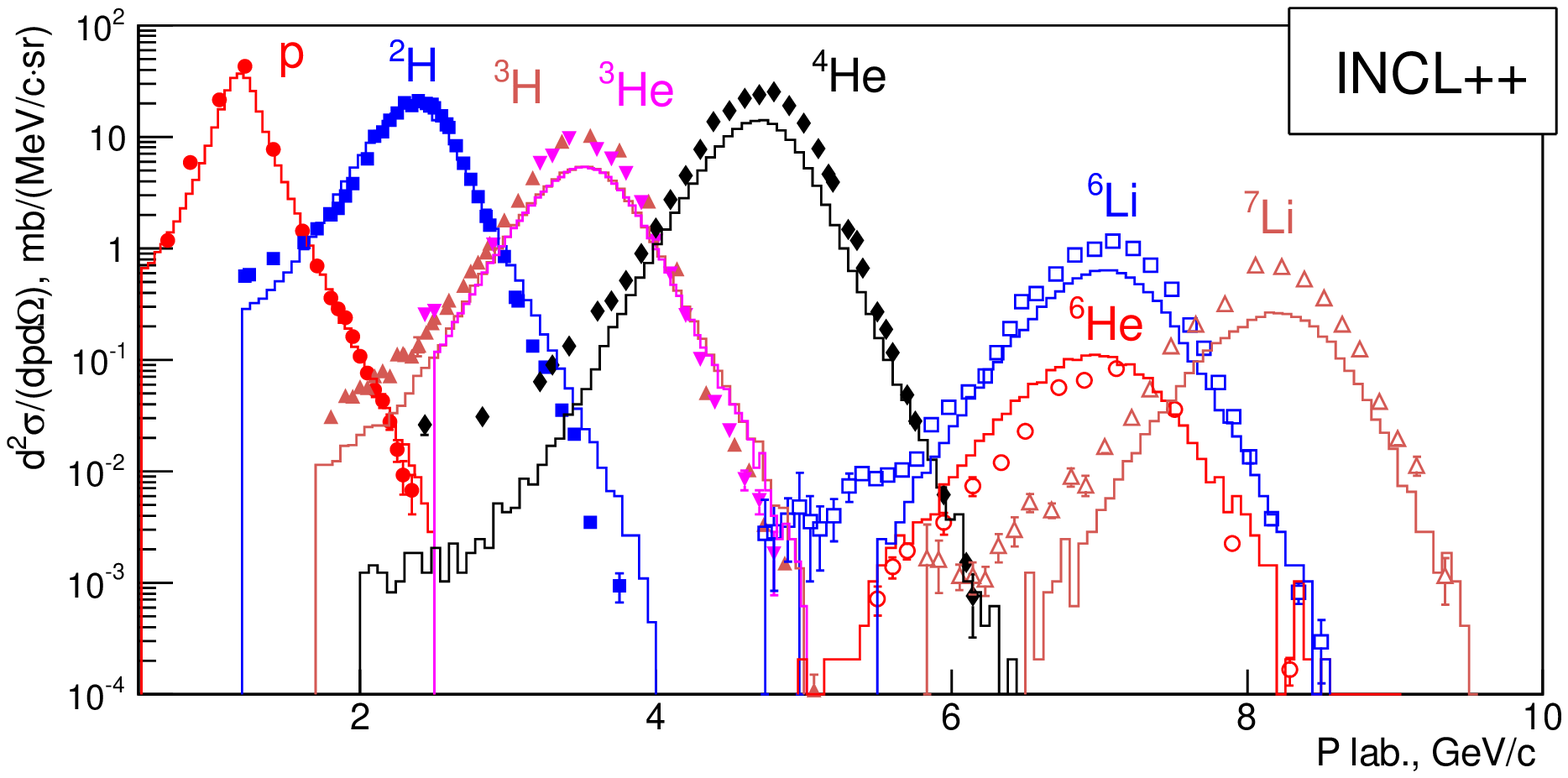}
}
        \subfigure[]{%
    \label{600MeVBC}
          \includegraphics[width=0.50\textwidth]{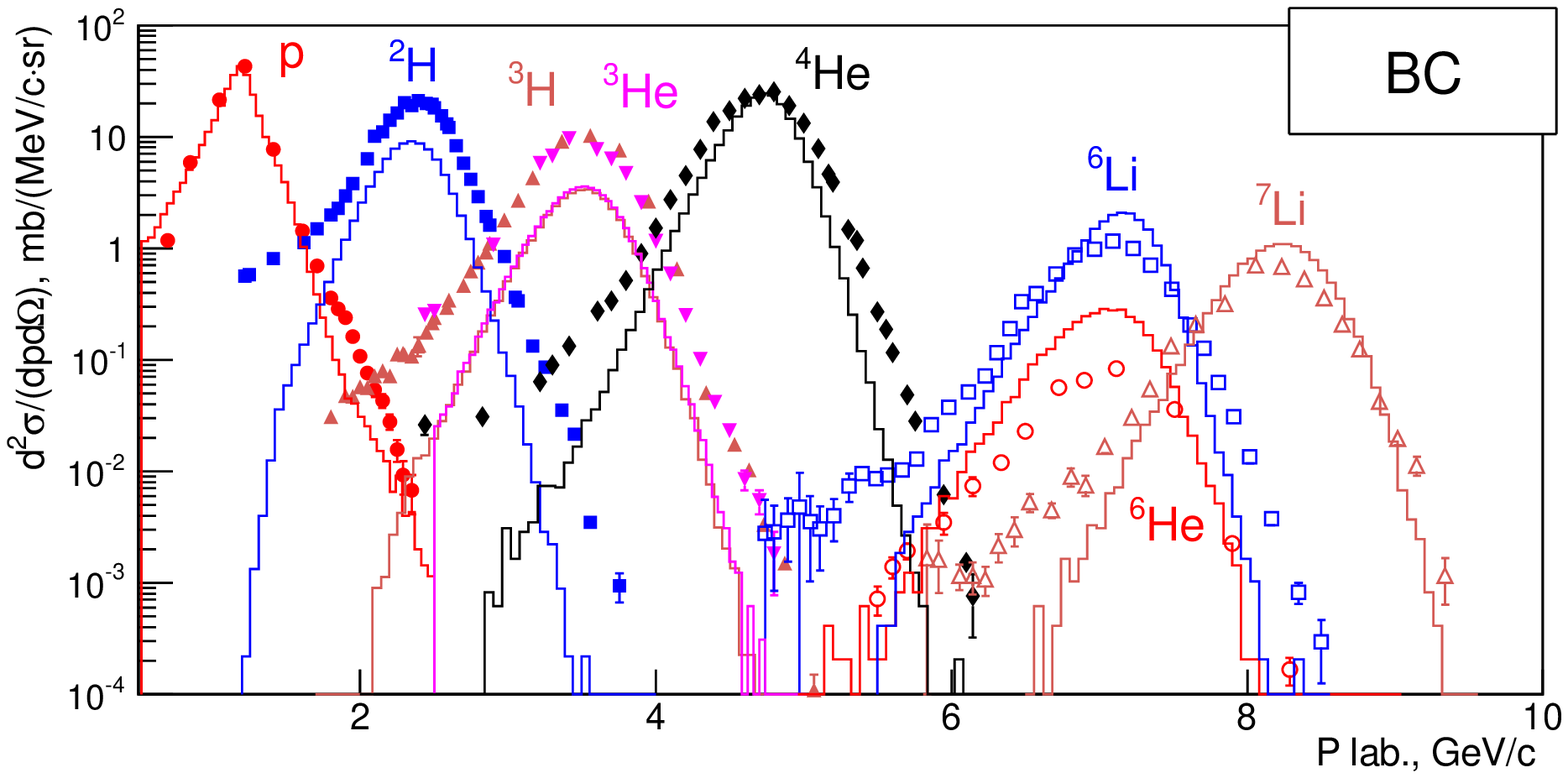}\hspace{-0.4cm}
}
        \subfigure[]{%
    \label{600MeVLA}\hspace{-0.3cm}
          \includegraphics[width=0.50\textwidth]{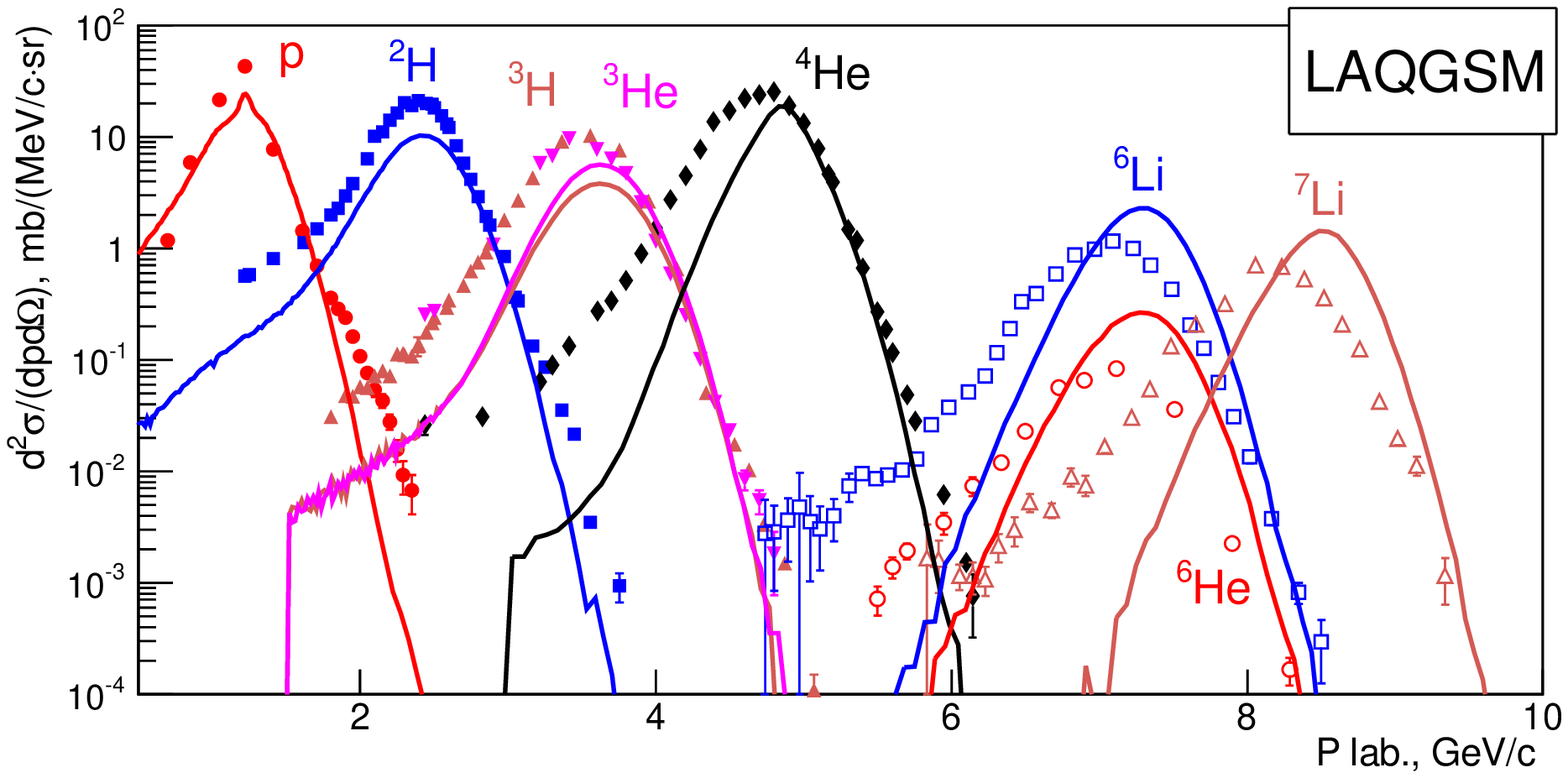}
}
        \subfigure[]{%
    \label{600MeVQMD}
          \includegraphics[width=0.50\textwidth]{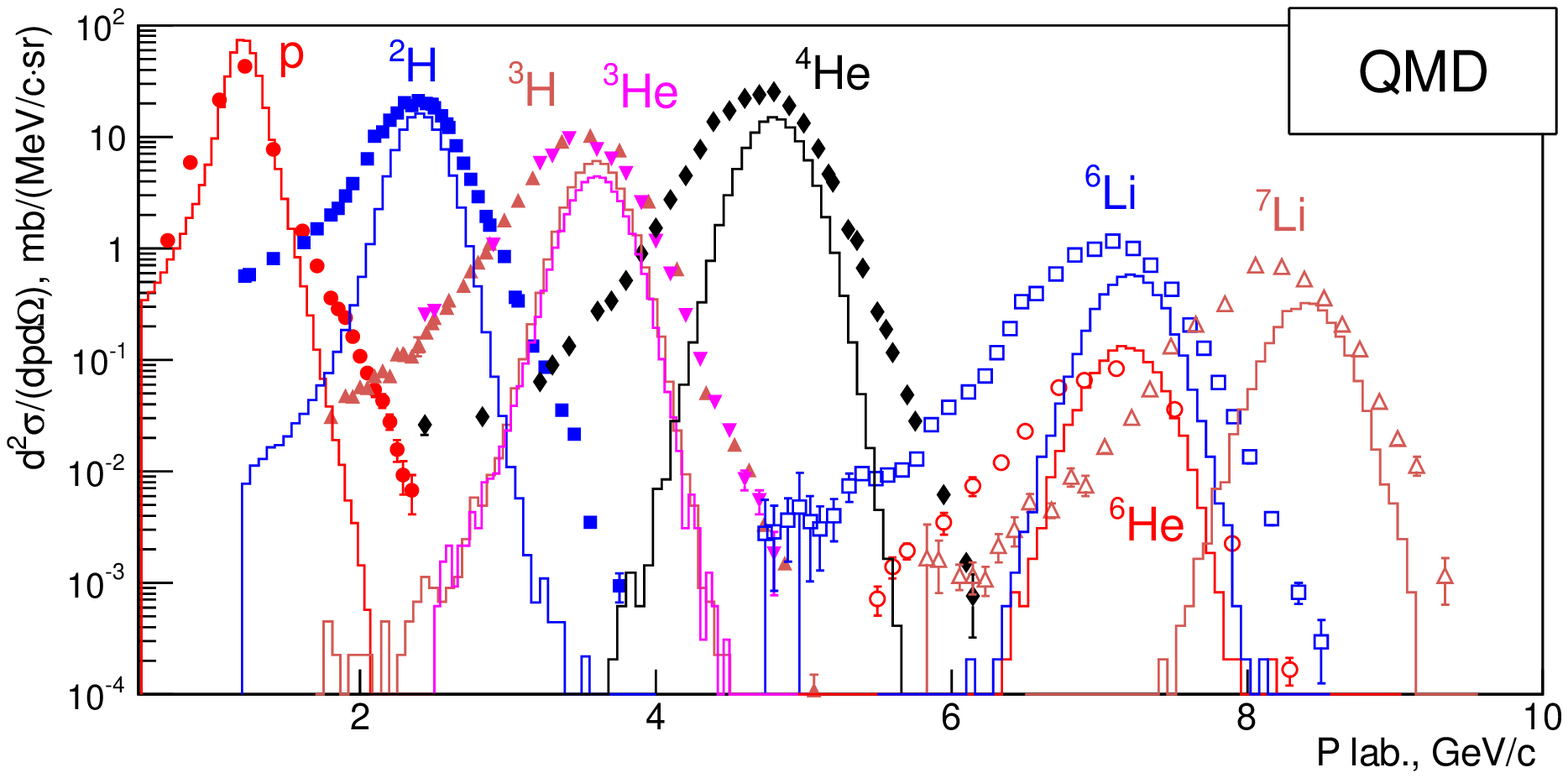}\hspace{-0.4cm}
}
     \end{center}
    \caption{%
Relative yields of H, He and Li isotopes in $^{12}$C + $^{9}$Be
interaction at 0.6 GeV/nucleon and at 3.5$^{\circ}$ as a functions of
fragment laboratory momenta. Data (points), model calculations
(histograms/lines): (a) INCL++ model \cite{dudouet}, (b) BC model
\cite{BC}, (c) LAQGSM model \cite{LAQGSM}, (d) QMD model
\cite{QMD}.
     }%
   \label{600MeVall}
\end{figure}

In the FRAGM experiment at ITEP TWA heavy ion accelerator, we have
measured the fragment  yields from the reaction
\begin{equation}
^{12}\mathrm{C} + ^{9}\hspace{-0.08cm}\mathrm{Be} \rightarrow \mathrm{f} + X
\end{equation}
with a beamline spectrometer set at 3.5$^{\circ}$ to carbon beam. Here
f stands for all fragments from protons up to isotopes of projectile
nucleus. The projectile kinetic energies were $T_{0}$= 0.2$-$3.2 GeV/nucleon.
In this report we present data at $T_{0}$= 0.6 GeV/nucleon for fragments:
hydrogen, helium and two lithium ($^{6}$Li, $^{7}$Li) isotopes.
The fragments were measured at a wide momentum region which include
the midrapidity, the fragmentation peak and the cumulative regions.
In the last one the fragment momenta per nucleon are much higher
than momentum per nucleon of the projectile. This gives a good testing
ground for a comparison with predictions of different ion-ion interaction
models.\\ The fragment yields were measured by scanning the beamline
spectrometer momentum with a step of 50$-$100 MeV/c and counting the
number of events corresponding to different fragments  and normalizing
to the monitor. The fragments were well separated on time-of-flight
$vs$ dE/dx plots. The relative cross sections $d^{2}\sigma/(d\Omega dp)$,
where $p$ is the fragment momentum in a laboratory frame, were calculated.
They are shown for hydrogen, helium and lithium isotopes in comparison
with the calculations by four models: INCL++ (Fig. \ref{600MeVINCL}), BC
(Fig. \ref{600MeVBC}), LAQGSM  (Fig. \ref{600MeVLA}) and QMD
(Fig. \ref{600MeVQMD}). We used the INCL++, BC and QMD models from a
GEANT4-package.
Our measurements cover three-to-six orders in the cross section
magnitude, depending on the fragment. Qualitatively, all the models
reproduce well the energy dependence of the differential cross
sections of the fragment yields. Model prediction for the cross
section at fragmentation peak maxima differ by no more than 2$-$3
times\footnote{The data of the FRAGM experiment were normalized to BC
calculations at the maximum of the proton peak. This normalization
factor was used in all figures.}. The largest differences are
observed for the QMD model, which predicts smaller width for the
fragmentation peaks. LAQGSM reproduces the energy dependence of
the cross sections at high energy part of the fragmentation peaks,
but underestimates the cross section in the low-energy part. The
predictions of BC and INCL++ models are very close, but INCL++
gives better description of the experimental data, which is
especially noticeable in the areas far from the fragmentation peak
maxima.
\section{Slope parameters from kinetic energy spectra}
\begin{figure}[ht!]
     \begin{center}
           \includegraphics[width=0.95\textwidth]{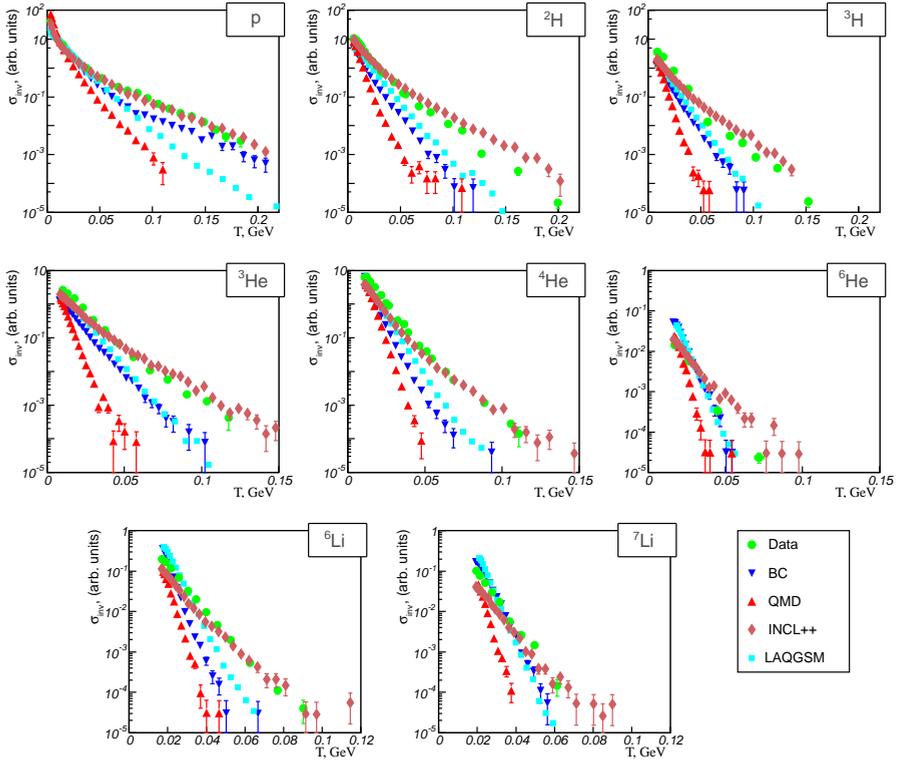}
     \end{center}
    \caption{%
Invariant cross sections as functions of fragment kinetic energies
in the $^{12}$C rest frame: measured data $vs$ model calculations.
     }%
   \label{600MeVinvTmax}
\end{figure}

\begin{figure}[ht!]
     \begin{center}
           \includegraphics[width=0.99\textwidth]{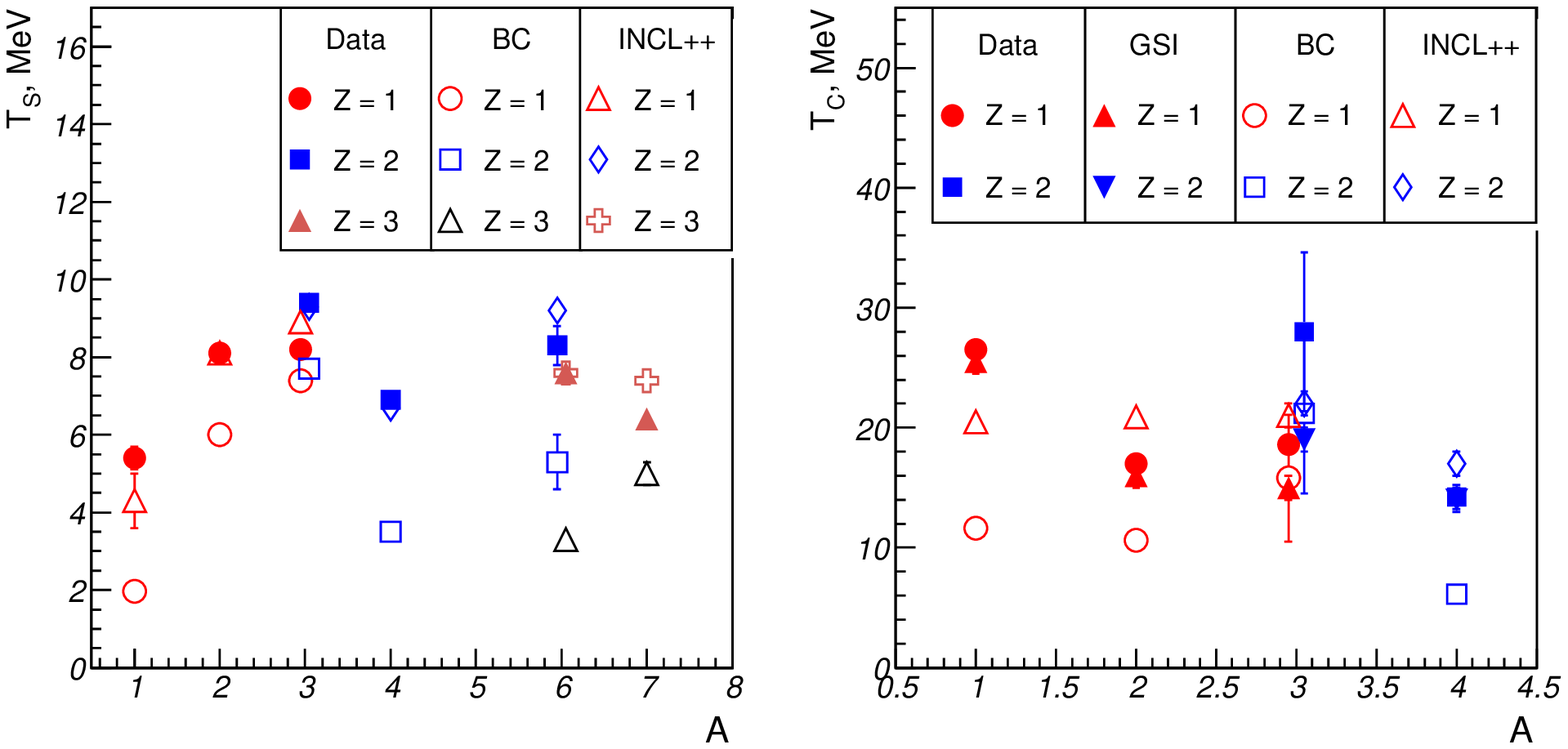}
     \end{center}
    \caption{%
Temperature parameters $T_{S}$ (left) and $T_{C}$ (right) as a function
of fragment atomic number. Closed points are ours, and GSI results are
from \cite{odeh}, open points are calculations in INCL++ and BC models.}%
   \label{600MeVTtable}
\end{figure}

In the framework of thermodynamical picture of nuclear fragmentation,
the fragment kinetic energy ($T$) distribution in the rest frame of
the carbon nucleus should be of Maxwell-Boltzmann type and not
depend on the fragment type. The distributions of invariant cross
section for fragment yields $Ed^3\sigma /d ^3
p = (E/p^2) d^{2}\sigma/(d\Omega dp)$, where $E$ is a total energy,
are shown in Fig. \ref{600MeVinvTmax} as function of $T$. Both
experimental data and model calculations are presented. The INCL++
model gives a good description of the experimental data much better
than the others. The spectra were parameterized by a sum of two
Maxwell-Boltzmann distributions
\begin{equation}
E d^3\sigma /d^3 p = E (A_S exp (- T / T_S) + A_C exp (- T /
T_C)),
\end{equation}
where $A_S$ and $A_C$ are normalization factors for low and high energy
regions, and the slope parameters $T_S$ and $T_C$ are "temperatures"
defined in these regions.  The function (2) gives a good description
of both the data and the calculations for all models and all fragments.
The obtained values of $T_S$ and $T_{C}$ are shown in
Fig. \ref{600MeVTtable}. The $T_S$ values are in a reasonable agreement
with those obtained in \cite{greiner} for 1$-$2 GeV/nucleon carbon ions.
The experimental results for $T_C$ from \cite{odeh} obtained at GSI at
1 GeV/nucleon for $^{197}$Au + $^{197}$Au collisions are in a reasonable
agreement with our results. The INCL++ model describes the data on $T_S$
well and better than the other models.
\section{Conclusion}
Fragment yields from the reaction $^{9}$Be($^{12}$C, f)X (f $-$ fragments
from p to $^{7}$Li) at  0.6 GeV/nucleon  were measured and compared to
prediction of four models of ion-ion interactions. The INCL++ describes
all momentum spectra rather well, both in the region of fragmentation
peak and in the cumulative region while all other models underestimate
the experimental results in the cumulative (high momentum) region.
Kinetic energy spectra in the projectile rest frame can be parameterized as
$E(A_{S}\exp(-T/T_{S})+A_{C}\exp(-T/T_{C}))$, where both $T_{S}$ and
$T_{C}$ values are in satisfactory  agreement with the predictions of the
INCL++ model. Other models strongly underestimates the data at high
kinetic energies;  $T_{C}$ values are higher for protons than for other
fragments.

Authors would like to thank I.I. Tsukerman for help. We are also
indebted to the personnel of TWAC-ITEP and technical staff of the
FRAGM experiment. The work has been supported in part by the RFBR
(grant No. 15-02-06308). Part of the work performed at LANL by S.G.M.
was carried out under the auspices of the National Nuclear Security
Administration of the U.S. Department of Energy at Los Alamos
National Laboratory under Contract No. DE-AC52-06NA25396.

\end{document}